\documentclass[conference]{IEEEtran}
\IEEEoverridecommandlockouts
\usepackage{./spconf}

\usepackage{amsmath,amssymb,amsthm}
\usepackage{enumerate}
\usepackage{stfloats}
\usepackage{comment}
\usepackage{subcaption}
\usepackage{siunitx}
\usepackage{mathtools}
\usepackage{accents,latexsym,cancel}
\usepackage{cite}

\usepackage[dvipsnames]{xcolor}
\definecolor{myPink}{RGB}{255,105,183}

\usepackage[T1]{fontenc}
\usepackage{graphics} 
\usepackage{epsfig} 
\usepackage[mathscr]{euscript}
\usepackage{algorithm}
\usepackage[noend]{algpseudocode}
\usepackage{bbm}
\makeatletter
\def\BState{\State\hskip-\ALG@thistlm}
\makeatother

\usepackage{tikz}
\usetikzlibrary{arrows,shapes,chains,matrix,positioning,scopes,patterns,calc,
decorations.markings,
decorations.pathmorphing,
}

\usepackage{pgfplots}
\pgfplotsset{compat=1.3}
\usepgflibrary{shapes}

\renewcommand{\epsilon}{\varepsilon}

\newcommand{\RNum}[1]{\uppercase\expandafter{\romannumeral #1\relax}}

\newcommand{\cv}{\ensuremath{\mathbf{c}}}

\newcommand{\mv}{\ensuremath{\mathbf{m}}}

\newcommand{\rv}{\ensuremath{\mathbf{r}}}
\newcommand{\sv}{\ensuremath{\mathbf{s}}}

\newcommand{\vv}{\ensuremath{\mathbf{v}}}

\newcommand{\wv}{\ensuremath{\mathbf{w}}}
\newcommand{\xv}{\ensuremath{\mathbf{x}}}
\newcommand{\yv}{\ensuremath{\mathbf{y}}}
\newcommand{\zv}{\ensuremath{\mathbf{z}}}

\newcommand{\zerov}{\ensuremath{\mathbf{0}}}

\newcommand{\etav}{\ensuremath{\boldsymbol{\eta}}}

\newcommand{\lambdav}{\ensuremath{\boldsymbol{\lambda}}}

\newcommand{\muv}{\ensuremath{\boldsymbol{\mu}}}

\newcommand{\zetav}{\ensuremath{\boldsymbol{\zeta}}}

\newcommand{\Phim}{\ensuremath{\boldsymbol{\Phi}}}
\newcommand{\Am}{\ensuremath{\mathbf{A}}}
\newcommand{\Dm}{\ensuremath{\mathbf{D}}}

\newcommand{\supp}{\ensuremath{\operatorname{supp}}}

\DeclareMathAlphabet{\mcl}{OMS}{cmsy}{m}{n}

\newlength\tikzwidth
\newlength\tikzheight

\definecolor{mycolor1}{rgb}{0.63529,0.07843,0.18431}%
\definecolor{mycolor2}{rgb}{0.00000,0.44706,0.74118}%
\definecolor{mycolor3}{rgb}{0.00000,0.49804,0.00000}%
\definecolor{mycolor4}{rgb}{0.87059,0.49020,0.00000}%
\definecolor{mycolor5}{rgb}{0.00000,0.44700,0.74100}%
\definecolor{mycolor6}{rgb}{0.74902,0.00000,0.74902}%

\title{A Hybrid Approach to Coded Compressed Sensing\\where Coupling Takes Place via the Outer Code}
\name{\begin{tabular}{c}Jamison R. Ebert,
Vamsi K. Amalladinne,
Jean-Francois Chamberland,
	Krishna R. Narayanan \end{tabular}
\thanks{
This material is based upon work supported, in part, by the National Science Foundation (NSF) under Grant CCF-1619085 and by Qualcomm Technologies, Inc., through their University Relations Program.}
}
\address{Department of Electrical and Computer Engineering, Texas A\&M University}

\begin{document}

\maketitle

\begin{abstract}
This article seeks to advance coded compressed sensing (CCS) as a practical scheme for unsourced random access.
The original CCS algorithm features a concatenated structure where an inner code is tasked with support recovery, and an outer tree code conducts message disambiguation.
Recently, a link between CCS and sparse regression codes was established, leading to the application of approximate message passing (AMP) to CCS.
This connection was subsequently strengthened by integrating AMP and belief propagation on the outer code through a dynamic denoiser.
Along these lines, this work shows how block diagonal sensing matrices akin to those used in traditional CCS, together with the aforementioned dynamic denoiser, form an effective means to get good performance at low-complexity.
This novel architecture can be used to scale this scheme to dimensions that were previously impractical.
Findings are supported by numerical simulations.
\end{abstract}


\section{Introduction}

Recently, there has been a renewed interest in research problems related to uncoordinated multiple access communications and sparse support recovery in exceedingly large dimensions, with a focus on the design of efficient low-complexity algorithms.
This attention stems, partly, from the introduction of unsourced random access as a paradigm to enable machine-driven wireless data transfers at scale, a prime application for future wireless infrastructures.
The aforementioned problems fall within the area of compressed sensing (CS), a topic that has seen significant theoretical and practical advances within the past decades.
In particular, many efficient solvers are readily available to perform sparse recovery, with both performance guarantees and low complexity.
Nevertheless, an important limitation in seeking to apply standard CS solvers to support recovery in exceedingly large dimensions comes from handling the ensuing sensing matrices.
For example, while it may be possible for such algorithms to handle sparse vectors with a million entries, computational limitations preclude the straightforward application of these solvers to much longer vectors.
This contrasts with the dimensionality of unsourced random access problems, which can easily exceed $2^{100}$.
Many schemes have been introduced to address this issue~\cite{ordentlich2017low, vem2019user, amalladinne2019coded, calderbank2018chirrup, fengler2019sparcs, pradhan2019sparseidma, marshakov2019polar, AKPolar, amalladinne2020enhanced, Amalladinne2020AMP}, each providing a way to circumvent the curse of large dimensions.

Coded compressed sensing (CCS), a line of investigation introduced by Amalladinne et al., has inspired several low-complexity schemes for unsourced random access.
This framework is rooted in a divide-and-conquer approach where support recovery is broken down into several sub-problems, each of a size amenable to the application of standard CS solvers, such as non-negative least squares (NNLS) or approximate message passing (AMP).
This reduction is enabled through an architecture that contains a concatenated code structure reminiscent of sparse regression codes~\cite{CIT-092} and for-all sparse recovery~\cite{gilbert2017all}.
Once fragments are obtained by the CS solvers, they are stitched together using the outer tree code, yielding the desired support of the sparse vector.
A significant advancement to this paradigm was proposed by Fengler, Jung, and Caire~\cite{fengler2019sparcs}.
They realized that the complexity reduction afforded by the CCS scheme originates primarily from the tree encoding and the corresponding partitioning of codewords.
Consequently, one can run AMP with a dense sensing matrix applied to a tree encoded sparse vector, rather than using individual CS solvers, thereby resulting in notable performance improvements.

The connection between coded compressed sensing and AMP can be pushed further.
In the original implementation~\cite{fengler2019sparcs}, the inner code is decoded first using AMP paired with a separable denoiser that accounts for the sparse nature of the input vector.
Once this phase is complete, the outer tree code is employed to piece fragments together.
In~\cite{Amalladinne2020AMP}, Amalladinne et al.\ show that the tree code can be redesigned to allow information to flow dynamically between the inner and outer decoders.
The resulting AMP denoiser takes advantage of both sparsity and the structure of the outer tree code.
This enhancement improves performance significantly while maintaining low overall complexity.

The present article seeks to advance the state-of-the-art in CCS via the following contributions.
We show that, in specific cases, the coupling afforded by the outer code is enough to capture most of the performance gains associated with CCS-AMP architectures.
That is, one can utilize a block diagonal sensing matrix in CCS-AMP, rather than a dense matrix, without incurring significant performance loss.
This can be interpreted as either a means to reduce complexity or a pathway to apply this algorithm to much larger systems, while maintaining the benefits of CCS-AMP.
We also note that the framework proposed in~\cite{Amalladinne2020AMP} works for a general class of graph-based codes suited to belief propagation, beyond the modified tree code introduced therein.
Thus, the proposed framework also raises an important research question: \emph{What are good outer codes for CCS-AMP?}

\section{System Model and Framework}

Mathematically, we wish to address a version of the noisy sparse support recovery problem, $\yv = \Phim \xv + \zv$ where $\xv$ is a $K$-sparse vector and $\zv$ is additive Gaussian noise with independent $\mathcal{N}(0,1)$ elements.
The design of the sampling matrix $\Phim$ is under our control, and its columns can be viewed as a dictionary of possible signals.
Our goal is to recover the support of $\xv$ from $\yv$ or, equivalently, to produce a binary vector $\hat{\xv}$ whose entries indicate the estimated support of $\xv$.
We assess performance based on the optimization criterion
\begin{equation*}
\textstyle
\frac{1}{K} \sum_{k \in \supp(\xv)} \left| 1 - \hat{\xv}(k) \right| .
\end{equation*}
The estimate $\hat{\xv}$ is also constrained to be (at most) $K$-sparse.

The model above is an instance of a compressed sensing problem, a well-studied object.
Our challenge is to design a pragmatic, low-complexity scheme for this particular problem when the dimensionality of $\xv$ precludes the direct application of existing CS solvers.
The motivation behind this setting comes from the unsourced multiple access problem formulation~\cite{polyanskiy2017perspective}, where the length of $\xv$ can easily exceed $2^{100}$.
Another intrinsic aspect of our formulation is its linear structure.
In unsourced random access, $\xv$ is obtained through the inherent addition that occurs on a multiple access channel.
In this sense, $\Phim \xv$ captures the summation of the signals that occurs with uncoordinated devices.
A detailed description of unsourced random access can be found in~\cite{polyanskiy2017perspective}.

Our envisioned architecture can be summarized by looking at three components.
First, a sequence of information bits is encoded into a signal.
This is followed by the composition of signal aggregates on a multiple access channel, plus the corruption resulting from additive noise.
Finally, a decoding process seeks to recover the list of transmitted messages by producing estimate $\hat{\xv}$.
It is instructive to keep in mind that the proposed scheme produces both a sampling matrix $\Phim$ and a support recovery algorithm, although the construction of the signal dictionary is only treated implicitly below.

\subsection{Encoding Procedure}

\begin{figure}[tbh]
  \centering
  \begin{tikzpicture}
  [
  font=\small, >=stealth', line width=1pt, draw=black,
  check/.style={rectangle, minimum height=6mm, minimum width=6mm, draw=black, fill=gray!20},
  section/.style={circle, minimum size=5mm, draw=black}
  ]

\node[rotate=90] (message) at (-2.75,0) {Message $\wv$};
\draw[->] (message.east) to [out=90,in=135] (-1.25,1.5);

\foreach \l in {1,2} {
  \node[section] (s\l) at (0.5,2.3-0.9*\l) {$s_{\l}$};
  \node (v\l) at (1.4,2.3-0.9*\l) {$\vv(\l)$};
}
\node (s4) at (0.5,-0.2) {$\vdots$};
\node (v4) at (1.4,-0.2) {$\vdots$};
\node[section] (sL) at (0.5,-1) {$s_{L}$};
\node (vL) at (1.4,-1) {$\vv(L)$};

\node[check] (a1) at (-1.5,0.9) {$a_1$};
\node[check] (ap) at (-1.5,-0.5) {$a_p$};
\node (a2) at (-1.5,0.3) {$\vdots$};

\node (variable) at (-0.5,-1.8) {Outer code};
\node (sparc) at (4.3,-1.8) {$\mv$};
\node (index) at (2.5,-1.8) {Indexing};

\draw (s1) -- (a1.east);
\draw (s2) -- (a1.east);
\draw (sL) -- (a1.east);
\draw (s2) -- (ap.east);
\draw (sL) -- (ap.east);

\foreach \l in {1,2} {
  \draw[draw=black] (4.2,2.75-0.9*\l) rectangle (4.4,1.85-0.9*\l);
  \node (m\l) at (3.6,2.3-0.9*\l) {$\mv(\l)$};
}
\node (d4) at (3.6,-0.2) {$\vdots$};
\draw[draw=black] (4.2,3.95-0.9*5) rectangle (4.4,3.05-0.9*5);
\node (mL) at (3.6,-1) {$\mv(L)$};

\draw[->] (v1) -- (m1);
\draw[->] (v2) -- (m2);
\draw[->] (vL) -- (mL);

\end{tikzpicture}
  \caption{
  Message $\wv$ is encoded into codeword $\vv = \vv(1) \cdots \vv(L)$ using a graph-based code. Every symbol $\vv(\ell)$ is translated into its index representation $\mv(\ell)$. These one-sparse blocks are concatenated, leading to vector $\mv$.}
  \label{figure:architecture1}
\end{figure}
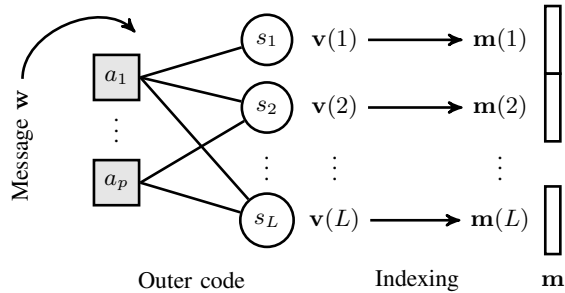

Our approach features a concatenated code, and it is depicted in Fig.~\ref{figure:architecture1}.
The outer code is assumed to come from a code family that is amenable to decoding through belief propagation~\cite{kschischang2001factor}.
For the problem at hand, this code should be over a large field or ring.
Together, these properties ensure that decoding can be performed efficiently using a combination of message passing and fast transform techniques~\cite{Amalladinne2020AMP}.
In the next step, the value of every variable node is translated into an index representation of length $m$.
This action is emblematic of CCS~\cite{amalladinne2019coded}, and it produces a one-sparse block.
The index vectors from the $L$ variable nodes are then aggregated into vector $\mv$, which possesses a structure reminiscent of sparse regression codes~\cite{CIT-092,rush2017capacity}.
The transmitted codeword is obtained by multiplying $\mv$ by a judiciously designed matrix $\Am \Dm$, i.e., $\Am \Dm \mv$ is transmitted over the channel.

\subsection{Decoding Strategy}

Within the proposed framework, the observation is composed of $K$ signals embedded in noise.
We must therefore extend our notation to accommodate several messages being encoded separately.
The observation, which acts as input to the decoding algorithm, becomes
\begin{equation*}
\textstyle
\yv = \sum_{i = 1}^K \Am \Dm \mv_i + \zv
= \Am \sv + \zv
\end{equation*}
where $\sv$ is the aggregate signal, $\Dm$ is a diagonal matrix that accounts for symbol power, and $\zv$ is noise.
The goal of the decoder is twofold: it must recover the support of $\sv$ from observation $\yv$, and it must disambiguate the list of codewords $\left\{ \mv_i \right\}$ that gives rise to $\sv$.

An algorithmic structure suitable to recover $\sv$ from $\yv$, as introduced in~\cite{fengler2019sparcs}, is the application of AMP.
The approach is to utilize AMP with a posterior mean estimate (PME) as the denoiser to first recover $\sv$, and then stitch fragments using the tree outer code.
Building on this contribution, it is shown in~\cite{Amalladinne2020AMP} that the outer code and the corresponding decoder can be redesigned to act, together with a PME, as a dynamic denoiser.
In this latter case, one round of belief propagation is performed on the factor graph of the outer code to inform the prior probabilities of the PME at every step.
Interestingly, the framework of~\cite{Amalladinne2020AMP} can be generalized to outer codes beyond tree codes.
This is meaningful because the extra flexibility acts as a means to circumvent the systematic encoding intrinsic to tree coding and, hence, it offers a richer design space for CCS.

Our main contribution comes from examining sensing matrix $\Am$.
The original CCS framework of Amalladinne et al.~\cite{amalladinne2019coded} can be interpreted as having a block diagonal structure with independent CS solvers, whereas CCS with AMP is applied to a dense matrix $\Am$~\cite{fengler2019sparcs}.
However, our new framework is rich enough to enable running an AMP-inspired algorithm with a block diagonal $\Am$, while performing the denoising jointly across blocks using one round of belief propagation on the factor graph of the outer code.
Conceptually, this hybrid scheme induces a form of spatial coupling through the joint denoising function while reducing the complexity of the inner decoder, providing excellent performance at a lower computational cost.
Alternatively, the hybrid approach can be leveraged to extend the proposed techniques to support recovery problems of much larger dimensions, beyond what was practically possible with existing solutions.
This should become manifest shortly, as we turn to the technical aspects of our scheme.

\section{AMP with Dynamic Denoiser}

We turn to the detailed description of the proposed framework.
We adopt an AMP-inspired composite iterative structure.
The algorithm iterates between two equations:
\begin{gather}
\zv^{(t)} = \yv - \Am \Dm \sv^{(t)} + \frac{\zv^{(t-1)}}{n} \operatorname{div} \Dm \etav_{t-1} \left( \rv^{(t-1)} \right) \label{equation:AMP-residual} \\
\sv^{(t+1)} = \etav_t \left( \Am^{\mathrm{T}} \zv^{(t)} + \Dm \sv^{(t)} \right)
 \label{equation:AMP-denoising}
\end{gather}
where $n$ denotes the height of the sensing matrix $\Am$ with initial conditions $\sv^{(0)} = \zerov$ and $\zv^{(0)} = \yv$.
Equation \eqref{equation:AMP-residual} can be interpreted as computing the \emph{residual} enhanced with an Onsager correction~\cite{bayati2011dynamics,donoho2013information};
whereas \eqref{equation:AMP-denoising} is a state update through denoising.
The argument to the denoising function $\etav_t (\cdot)$ is termed the \emph{effective observation}, and it is denoted by $\rv^{(t)}$.
When $\Am$ is dense, this algorithm falls within the extended AMP framework for non-separable denoisers characterized by Berthier, Montanari, and Nguyen~\cite{berthier2017state}.
A remarkable fact about AMP is that, under certain technical conditions, the effective observation is asymptotically distributed as $\mathbf{D} \sv + \tau_t \zetav_t$ where $\zetav_t$ is an i.i.d.\ $\mathcal{N}(0,1)$ random vector and $\tau_t$ is a deterministic quantity.
This property, which hinges on the presence of the Onsager correction in \eqref{equation:AMP-residual}, is pertinent because it permits good intuition and makes AMP mathematically tractable in many contexts.

The purpose of the denoising function is to incorporate knowledge about properties of $\sv$ into the composite iterative process.
By construction, $\sv$ and $\rv$ inherit a block representation akin to $\mv$, as illustrated in Fig.~\ref{figure:architecture1}.
Under this block representation, $\sv$ features $L$ sections, each with $K$ non-zero entries
(with high probability for large sections).
This is captured in the original PME denoiser.
A less obvious property of $\sv$, and one that is more challenging to leverage, comes from the fact that it arises as a sum of $K$ valid codewords from the outer code.
Consequently, it must decompose into individual messages $\mv_i$ that are consistent with the factor graph of the outer code.
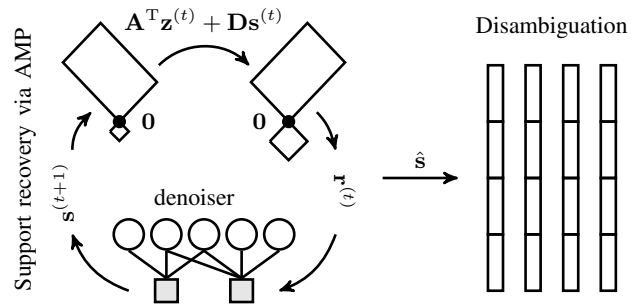
\begin{figure}[t!]
  \centering
  \begin{tikzpicture}[
  font=\small, >=stealth', line width=1pt,
  check/.style={rectangle, minimum height=3mm, minimum width=3mm, draw=black, fill=gray!20},
  section/.style={circle, minimum size=4mm, draw=black},
  zero/.style={circle, minimum size=0.2mm, draw=black, fill=black}
]

\node[rotate=90] (update) at (-0.75,-0.9375) {$\sv^{(t+1)}$};
\draw[->] (-0.625,-0.375) to [out=90,in=225] (-0.375,0.125);
\draw[->] (0.125,-2.25) to [out=160,in=-70] (-0.625,-1.5);

\node[rotate=-90] (Statistic) at (3,-0.9375) {$\rv^{(t)}$};
\draw[<-] (2.875,-0.375) to [out=105,in=-45] (2.625,0.125);
\draw[<-] (2.125,-2.25) to [out=20,in=-110] (2.875,-1.5);

\filldraw (0,0) circle (2pt) node[right,xshift=1ex] {$\mathbf{0}$};
\draw[-] (-0.125,-0.13125) -- (0,-0.25625) -- (0.125,-0.13125) -- (-0.75,0.7875) -- (-0.25,1.3125) -- (0.5,0.525) -- (-0.125,-0.13125);

\foreach \x in {2.25}
{
\filldraw (\x,0) circle (2pt) node[left,xshift=-1ex] {$\mathbf{0}$};
\draw[-] (-0.25+\x,-0.2625) -- (0+\x,-0.5125) -- (0.25+\x,-0.2625) -- (-0.5+\x,0.525) -- (0.25+\x,1.3125) -- (0.75+\x,0.7875) -- (-0.25+\x,-0.2625);
}

\draw[->] (0.525,0.75) to [out=45,in=135] (1.725,0.75);
\node (AT) at (1.125,1.35) {$\Am^{\mathrm{T}} \zv^{(t)} + \Dm \sv^{(t)}$};

\foreach \l in {1,2,3,4,5} {
  \node[section] (s\l) at (-0.375+0.5*\l,-1.5) {};
}
\node[check] (a1) at (0.625,-2.25) {};
\node[check] (a2) at (1.625,-2.25) {};
\draw (s1.south) -- (a1.north);
\draw (s2.south) -- (a1.north);
\draw (s3.south) -- (a1.north);
\draw (s3.south) -- (a2.north);
\draw (s2.south) -- (a2.north);
\draw (s4.south) -- (a2.north);
\draw (s5.south) -- (a2.north);
\node (denoiser) at (1,-1) {denoiser};

\foreach \l in {1,2,3,4} {
  \draw[draw=black] (4.9,1.5-0.75*\l) rectangle (5.1,0.75-0.75*\l);
  \draw[draw=black] (5.4,1.5-0.75*\l) rectangle (5.6,0.75-0.75*\l);
  \draw[draw=black] (5.9,1.5-0.75*\l) rectangle (6.1,0.75-0.75*\l);
  \draw[draw=black] (6.4,1.5-0.75*\l) rectangle (6.6,0.75-0.75*\l);
}

\node[rotate=90] (recovery) at (-1.25,-0.5) {Support recovery via AMP};
\node (disambiguation) at (5.75,1.25) {Disambiguation};
\draw[->] (3.5,-0.75) to node[above] (shat) {$\hat{\sv}$} (4.5,-0.75);

\end{tikzpicture}
  \caption{The envisioned decoder features two phases.
  Initially, AMP is employed to estimate the support of $\sv$ through a composite iteration scheme with a BP denoiser
  This is followed by a disambiguation phase where individual codewords are recovered from the aggregate $\hat{\sv}$.
  }
  \label{figure:architecture2}
\end{figure}

Describing the intricacies of the BP denoiser is beyond the scope of this article, but this information can be found in~\cite{amalladinne2020unsourced}.
Essentially, a bipartite factor graph is used to compute the beliefs that coded symbols from a specific device take on certain values.
Messages are exchanged locally and they are based on extrinsic information.
This operation parallels the evolution of single-user decoders~\cite{kschischang2001factor}, albeit with key modifications.
The local observations within the factor graph are given by the PME~\cite{fengler2019sparcs}, as a means to circumvent combinatorial complexity associated with the fact that several codewords exist on a same factor graph.
Section sizes must be large enough to prevent collisions with high probability and, consequently, the graph-based code typically employs a large alphabet.

Within the context of AMP, the denoiser seeds the factor graph with local observations, and then performs one round of belief propagation.
The information messages coming to a variable node, including that derived from the local observation, are combined as to produce a local state estimate.
These components are then aggregated to produced an updated state, which serves as input to the computation of a residual.
For our system, the Onsager term is $\big( \big\| \Dm^2 \sv^{(t)} \big\|_1 - \big\| \Dm \sv^{(t)} \big\|^2 \big)
/n \tau_{t-1}^2$.
We note that $\tau_t^2$ can be approximated as $\tau_t^2 \approx \left\| \zv^{(t)} \right\|^2/n$ for $t \ge 0$.
A justification for the same can be found in~\cite{SchniterAMP}.
Message passing over large sections can be performed efficiently using the fast Fourier transform or the Fast Walsh-Hadamard transform for a suitably designed outer code as in~\cite{Amalladinne2020AMP}.
Once the AMP algorithm has converged, the disambiguation process uses the structure of the outer code to stitch message fragments together, as in~\cite{Amalladinne2020AMP}.
A notional diagram appears in Fig.~\ref{figure:architecture2}.

\subsection{Candidate Implementations}


In this investigation, we compare the performance of competing implementations for CCS.
The first scheme is the original CCS scheme introduced in~\cite{amalladinne2019coded}.
The corresponding sensing matrix $\Am$ is block diagonal, and we use AMP as a CS solver on a per-section basis.
Recovered message fragments are stitched together using the tree code afterwards.
Candidate~2 is CCS-AMP, with its dense matrix $\Am$, as reported in~\cite{fengler2019sparcs}.
For this system, AMP is used as a global CS solver, followed by disambiguation using the tree code.
The third candidate implemented is CCS-AMP with dynamic denoising, whereby belief propagation is integrated into the AMP cycle~\cite{Amalladinne2020AMP}.
This scheme also employs a dense matrix~$\Am$.
Finally, the novel and forth implementation features a block diagonal $\Am$.
AMP is applied to individual blocks in that the residual is calculated concurrently, but section by section.
However, during the denoising phase, belief propagation is applied to the entire system to create coupling among blocks.
In the numerical evaluation, our proposed scheme corresponds to Case~4.

We note that the latter algorithm is a hybrid between CCS and CCS-AMP with dynamic denoising.
In this sense, it is an AMP-inspired framework that seeks to reap the benefits of a dense matrix at a much lower complexity.
It is worth noting that, although CCS-AMP is designed to handle very large dimensions, it is constrained in that the width of $\Am$ should not exceed $2^{20}$ or so.
In the hybrid implementation the state update is applied block-wise after denoising and, hence, the width of a block is constrained but not the width of the overall block diagonal sensing matrix.
The BP on the factor graph of the outer code, especially with a fast implementation, can be made to handle a large number of blocks.
This enables a scaling of the problem that was not possible in the original CCS because of error propagation due to hard decision, or CCS-AMP due to processing power.
This is also a distinguishing attribute of our novel scheme.

\section{Parameters and Numerical Results}

This section compares the performance of the proposed hybrid scheme against that of previously published CCS implementations. 
We consider a system with $K=100$ active devices.
The payload size corresponding to each active user is $w = 128$ bits.
The message of each such device is encoded into a block of $n=38400$ channel uses.
The \textit{energy-per-bit} for this system is defined as $\frac{E_{\mathrm{b}}}{N_0} = \frac{nP}{2w}$, where $P$ represents the energy of each symbol transmitted by the active users.
The factor graph for the outer code employed in our simulations is analogous to the triadic design introduced in~\cite{amalladinne2020unsourced}, and it is identical for all implementations.
Specifically, outer codewords are partitioned into $L=16$ blocks, each of length $16$~bits.
After switching to the index representation, section $\mv (\ell)$ has length $2^{16}$, yielding an overall vector $\mv$ of length $L 2^{16} = 2^{20}$.
Both the block diagonal matrix for CCS and CCS-hybrid, and the dense matrix for the two CCS-AMP variants are formed by selecting random rows from Hadamard matrices (excluding the row of all ones).
These design choices aim at providing a fair comparison for alternate systems.

\begin{figure}[t]
\centerline{

        
\begin{tikzpicture}
\definecolor{cBlue}{rgb}{0.2, 0.6, 1.0}  
\definecolor{cRed}{rgb}{1.0, 0.0, 0.0}   
\definecolor{cGreen}{rgb}{0, 0.6, 0}     
\definecolor{cBlack}{rgb}{0.0, 0.0, 0.0} 

\begin{semilogyaxis}[
font=\small,
width=6.5cm,
height=5cm,
scale only axis,
every outer x axis line/.append style={white!15!black},
every x tick label/.append style={font=\color{white!15!black}},
xmin=1.5,
xmax=4.5,
xtick={1.5, 2, 2.5,..., 4.5},
xlabel={$E_{\mathbf{b}}/N_0$},
xmajorgrids,
xminorgrids,
every outer y axis line/.append style={white!15!black},
every y tick label/.append style={font=\color{white!15!black}},
ymin=0,
ymax=1,
ytick={0.01, 0.1, 1.0},
ylabel={Per-User Error Rate $P_{\mathrm{e}}$},
ymajorgrids,
yminorgrids,
legend style={at={(0, 0)},anchor=south west,draw=black, fill=white, legend cell align=left,font=\footnotesize}
]


\addplot[color=cBlue, densely dotted, line width=2.0pt, mark size=1.4pt, mark=0, mark options={solid}]
table[row sep=crcr]{
1.5 0.9625 \\
1.75 0.9216 \\
2 0.8541 \\
2.25 0.7769 \\
2.5 0.6545 \\
2.75 0.5095 \\
3 0.381   \\
3.25 0.2481 \\
3.5 0.1651 \\
3.75 0.0966 \\
4  0.0552 \\
4.25 0.0295  \\
4.5 0.0149 \\
};
\addlegendentry{Case~1: CCS};

\addplot[color=cGreen, densely dashdotted, line width=2.0pt, mark size=1.4pt, mark=0, mark options={solid}]
table[row sep=crcr]{
1.5 0.9567\\
1.75  0.9204\\
2  0.8497 \\
2.25 0.7591 \\
2.5 0.6271 \\
2.75 0.4885 \\
3 0.3526 \\
3.25 0.2428 \\
3.5 0.1567 \\
3.75 0.0871 \\
4  0.0508 \\
4.25 0.0276 \\
4.5 0.0142 \\
};
\addlegendentry{Case~2: CCS-AMP w/o BP};

\addplot[color=cBlack, dashdotted, line width=2.0pt, mark size=1.4pt, mark=0, mark options={solid}]
table[row sep=crcr]{
1.5 0.68  \\
1.75 0.562  \\
2 0.433 \\
2.25  0.3023 \\
2.5 0.2097 \\
2.75 0.1322 \\
3 0.0862\\
3.25  0.0533 \\
3.5 0.0315 \\
3.75 0.0204 \\
4   0.0155 \\
4.25 0.0103\\
4.5 0.0075 \\
};
\addlegendentry{Case~3: CCS-AMP with BP};

\addplot[color=cRed, solid, line width=2.0pt, mark size=1.4pt, mark=0, mark options={solid}]
table[row sep=crcr]{
1.5 0.6934 \\
1.75 0.5825 \\
2 0.4498 \\
2.25 0.3279 \\
2.5 0.2237 \\
2.75 0.1502\\
3  0.0973 \\
3.25 0.0552 \\
3.5 0.0361 \\
3.75 0.0234\\
4  0.0172 \\
4.25 0.013  \\
4.5 0.0077  \\
};
\addlegendentry{Case~4: CCS-Hybrid};

\end{semilogyaxis}
\end{tikzpicture}}
  \caption{This graph offers a comparison of per-user error rate for four candidate AMP implementations.
  The performance of the coupled block-diagonal scheme is very competitive.}
  \label{fig:errorRatesK100}
\end{figure}
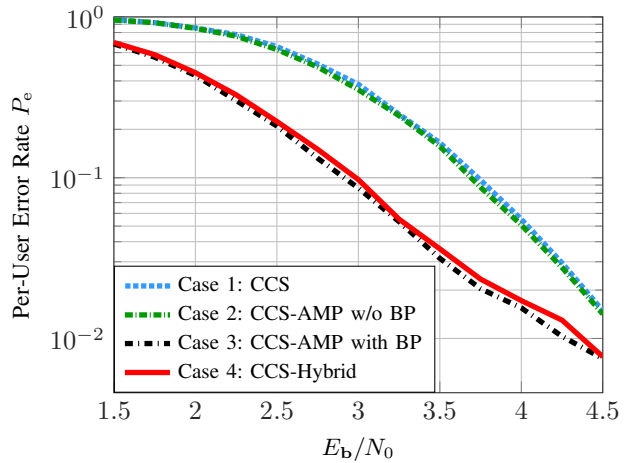
Numerical results appear in Fig.~\ref{fig:errorRatesK100}, where per-user probability of error is plotted as a function of $\frac{E_{\mathrm{b}}}{N_0}$.
Every point on this graph is averaged over 100 instances for statistical accuracy.
Interestingly, the performance of the hybrid scheme far exceeds those of the original CCS algorithm~\cite{amalladinne2018couple} or the original CCS-AMP~\cite{fengler2019sparcs}.
Furthermore, it approaches the performance of the more complex CCS-AMP implementation with the dynamic BP denoiser~\cite{Amalladinne2020AMP}.
This is very encouraging, as the type of spatial coupling generated by the outer code seems to be sufficient to guide the decoding closer to the true sparse solution.
Table~\ref{table:runTime} shows the relative computational loads of the four candidate schemes.
The hybrid CCS-AMP scheme considerably outperforms the CCS-AMP scheme without BP and admits a comparable execution time.

\begin{table}[bh]
\centerline{
\begin{tabular}{||c|c|c|c||}
\hline
Case 1 & Case 2 & Case 3 & Case 4 \tabularnewline
\hline
$1$ & $2.7369$ & $6.5078$ & $2.9028$ \tabularnewline
\hline
\end{tabular}
}
\caption{This table offers a run-time comparison between various candidate schemes normalized by the run-time of the original CCS scheme (Case 1).}
\label{table:runTime}
\end{table}

\section{Discussion}

This article explores a variation of coded compressed sensing (CCS), which adopts the simple block diagonal structure for the sensing matrix of the inner code but integrates message passing on the factor graph of the outer code as a means to produce spatial coupling.
The end result is a structurally simpler version of CCS that exhibits most of the performance benefits associated with the more complex CCS-AMP with belief propagation.
The complexity of the decoding process for the inner code becomes linear.
This enables the scaling of unsourced random access to problem sizes that could not be handled in the past, a situation which invites the search for better outer codes.

\newpage

\bibliographystyle{IEEEbib}
\bibliography{IEEEabrv,MACcollison}

\end{document}